\documentclass[%
 aip,
 cha,
 amsmath,amssymb,
 reprint,%
]{revtex4-1}

\usepackage{graphicx}
\usepackage{dcolumn}
\usepackage{bm}

\usepackage[utf8]{inputenc}
\usepackage[T1]{fontenc}
\usepackage{mathptmx}
\usepackage{etoolbox}
\usepackage{subfigure}
\makeatletter
\def\@email#1#2{%
 \endgroup
 \patchcmd{\titleblock@produce}
  {\frontmatter@RRAPformat}
  {\frontmatter@RRAPformat{\produce@RRAP{*#1\href{mailto:#2}{#2}}}\frontmatter@RRAPformat}
  {}{}
}%
\makeatother
\usepackage[usenames,dvipsnames,svgnames]{xcolor}

\begin{document}

\title{Tipping in an adaptive climate network model}
\author{T. Bdolach}
 \affiliation{Department of Mathematics, Humboldt University Berlin, 10099 Berlin, Germany}
\author{J. Kurths}%
\affiliation{ 
Potsdam Institute for Climate Impact Research, Member of the Leibniz Association, P.O Box 6012 03, D-14412 Potsdam, Germany}
\affiliation{Department of Physics, Humboldt University Berlin, 10099 Berlin, Germany}
\author{S. Yanchuk}
\affiliation{
School of Mathematical Sciences, University College Cork, Western Road, Cork T12 XF62, Ireland
}
\affiliation{
Potsdam Institute for Climate Impact Research, Member of the Leibniz Association, P.O Box 6012 03, D-14412 Potsdam, Germany}
\date{\today}

\begin{abstract}
With rising global temperatures Earth's tipping elements are becoming increasingly more vulnerable to crossing their critical thresholds. The reaching of such tipping points does not only impact other tipping elements through their connections, but can also have further effect on the global mean surface temperature (GMT) itself, either increasing or decreasing the probability of further tipping points being reached. Recently, a numerical study analyzing the risk of tipping cascades has been conducted, using a conceptual model describing the dynamics of a tipping element with its interactions with other tipping elements taken into account\cite{wunderling21}. Here, we extend the model substantially by including adaptation, so that the GMT-feedback induced by the crossing of a tipping point is incorporated as well. 
We find that although the adaptive mechanism does not impact the risk for the occurrence of tipping events, large tipping cascades are less probable due to the negative GMT-feedback of the ocean circulation systems. Furthermore, several tipping elements can play a different role in cascades in the adaptive model. In particular, the Amazon rainforest could be a trigger in a tipping cascade. Overall, the adaptation mechanism tends to slightly stabilize the network. 
\end{abstract}

\maketitle

\begin{quotation}
Tipping elements of the climate system are large-scale regions of the Earth that are sufficiently connected to be considered as a single dynamical unit that can exhibit threshold behavior (tipping). 
Such tipping can be a response to an increase in mean surface temperature, and can lead to significant climate changes and hence impacts on human life. Here we propose an extension of the model of interacting tipping elements of Wunderling et.~al.~\cite{wunderling21}, mainly by incorporating an adaptive feedback of the tipping elements on the global mean surface temperature (GMT), but also by including additional tipping elements. As a consequence, our model shows qualitative changes, such as shortening of the tipping cascades, different roles of some tipping elements, such as a possible increase in the role of the Amazon rainforest region in the tipping cascades.
\end{quotation}

\section{\label{sec:level1}Introduction}

Climate tipping elements are large-scale subsystems of the Earth system that can significantly alter their state once a critical value (tipping point) for a control parameter is exceeded. Such changes can be traced to an increase in global mean surface temperature (GMT) \cite{mckay22, lenton08}. Beyond this critical threshold, changes in the system become self-perpetuating and effectively irreversible due to hysteresis behavior \cite{mckay22}. In the case of global warming, it becomes more likely for tipping elements to cross their critical thresholds, which could have severe impacts on the climate system and the system path as a whole\cite{wunderling21, mckay22, steffen18}. Several studies even suggest that some tipping elements, mostly belonging to the Earth's cryosphere, are approaching or may have already reached their critical thresholds \cite{wunderling21,lenton19}.

Furthermore, the tipping elements can interact with each other. If the threshold of an element is crossed, it can raise or lower the probability that the threshold of another element is crossed as well
\cite{wunderling21, mckay22, wunderling24, kriegler09}. 
Destabilizing interaction links between pairs of tipping elements increase the probability for the occurrence of tipping cascades, i.e. chain reactions in which several tipping elements cross their critical thresholds, triggered by the tipping of a single element \cite{wunderling21, wunderling24, lenton19}. The feedbacks induced by a global cascade of tipping points could cause the entire Earth system to tip irreversibly into a significantly warmer so-called "Hothouse Earth" state \cite{steffen18}.

In Ref.~\onlinecite{wunderling21}, Wunderling \textit{et.~al.} conceptually modeled a network of four tipping elements in the Earth system - the Greenland ice sheet, the West Antarctic ice sheet, the Atlantic meridional overturning circulation (AMOC), and the Amazon rainforest - using a weighted dynamical network, with the individual dynamics of each element described by a normal form for a double-fold  bifurcation, extended by a linear coupling term to account for interactions between the tipping elements.

Although this approach seems to be an interesting model for the dynamics of tipping elements, it does not take into account the resulting feedback caused by the threshold crossing of the tipping elements \cite{mckay22,steffen18} on the GMT; the GMT is only considered as a parameter in Ref.~\onlinecite{wunderling21}.  
We therefore propose an adaptation mechanism in which the GMT depends on the states of the tipping elements.

Furthermore, several tipping elements characterized as global core tipping elements in Ref.~\onlinecite{mckay22} and their interactions are not considered in the model from Ref.~\onlinecite{wunderling21}. The Labrador-Irminger seas convection has been recently proposed as a tipping element \cite{mckay22}, \cite{swingedouw21} and it is expected to cross its threshold for global warming below 2 °C \cite{mckay22}. Although the East Antarctic ice sheet is not expected to reach a tipping point for a GMT-rise below 7.5 °C \cite{mckay22}, parts of it - the East Antarctic subglacial basins - have a significantly lower critical threshold temperature \cite{lenton19,mckay22}, and as a member of the Earth's cryosphere have clear destabilizing interactions with the other cryosphere elements \cite{mckay22}. We therefore extend the network of tipping elements to a six-element network including the Labrador-Irminger seas convection and the East Antarctic subglacial basins.

The structure of this paper is as follows: in Section 2, we recall the definition of a tipping element as presented in Refs.~\onlinecite{mckay22}~\&~\onlinecite{lenton08} as well as the tipping elements considered in the model. Furthermore, we describe the interactions of the two additional tipping elements within the network. In Section 3, we recall the model describing the dynamics within the network of tipping elements presented in Ref.~\onlinecite{wunderling21}. We then extend the model to an adaptive dynamical system in which the GMT depends on the states of the tipping elements. In Section 4, we perform the numerical study conducted in Ref.~\onlinecite{wunderling21} for the model with adaptation. A comparison of the results for the models with and without adaptation follows in Section 5. We summarize our results in Section 6.

\section{Tipping elements}
\subsection{The notion of tipping element}
We consider subsystems of the Earth system corresponding to a specific region and having  at least a subcontinental scale \cite{lenton08}.
In Ref. \onlinecite{lenton08}, such a subsystem is classified as a \textit{tipping element} if it is possible to combine the parameters controlling it into a single control $\rho$, for which there exists a critical value $\rho_{\text{crit}}$ such that any further increase of $\rho$ beyond $\rho_{\text{crit}}$ induces a qualitative change in the system.

It is assumed that there is a scalar system property $F$ of the tipping element, which is an indicator of the tipping process. 
For example, such a system feature for an ice sheet could be the volume of the ice \cite{lenton08}.
For the control $\rho$, one usually considers the rise in the GMT above pre-industrial levels (1850-1900). A more specific definition in which the GMT-rise takes the role of the control parameter is given in Ref. \onlinecite{mckay22}. According to this definition, a subsystem of the Earth system is classified as a tipping element if the following holds:\\
\textit{i.)} Changes in the subsystem become self-perpetuating once a critical value for GMT is reached, i.e. there exists a tipping point for this subsystem.\\
\textit{ii.)} These changes impact the Earth system significantly.\\
Once the tipping point is reached, the processes continue to unfold without any external forcing. Therefore, the impacts of reaching tipping points are usually irreversible, or effectively irreversible in human timescales \cite{mckay22}.

\begin{figure*}[htb]
 \includegraphics[scale=0.55]{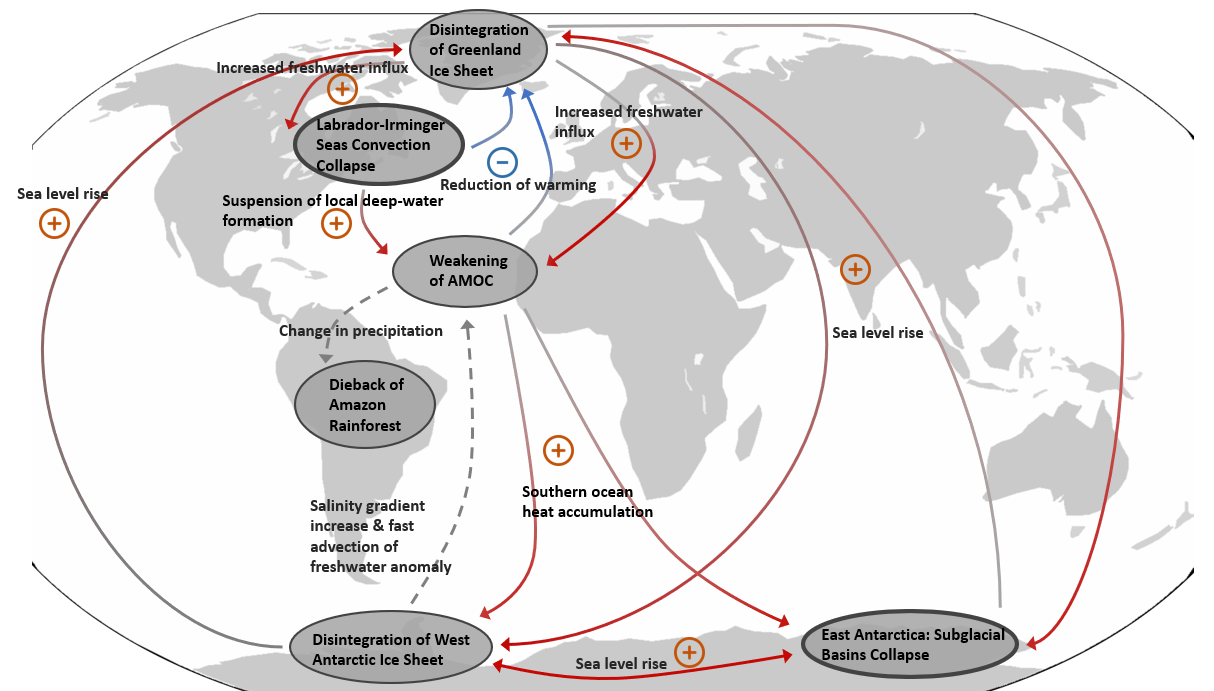}
 \caption{Extended network of six tipping elements and their interactions. The two tipping elements added into the model in this paper (compared to Ref.~\onlinecite{wunderling21}) have a thicker outline for clarity. The color-coding is the same as in Ref. \onlinecite{wunderling21}, with destabilizing interactions depicted in red, stabilizing interactions depicted in blue, and dashed lines denoting unclear interaction links.
 }
 \label{fig:mesh1}
 \end{figure*}

Tipping elements are generally not isolated from each other \cite{wunderling21, wunderling24, fan21}. To the set of proposed tipping elements belong cryosphere, biosphere and ocean circulation components of the Earth system, see Fig.~\ref{fig:mesh1}. If one of these elements crosses its threshold it can therefore influence other elements by the physical impacts of its tipping, e.g. sea-level rise, regional temperature and precipitation changes, etc. \cite{mckay22, kriegler09, wunderling24, fan21}. Such coupling between a pair of tipping elements could have a stabilizing or destabilizing effect \cite{wunderling21}. Thus, the tipping of a single element could decrease or increase the probability that another element would tip as well \cite{wunderling21}, possibly resulting in tipping cascades involving several tipping elements tipping after each other in a domino-like effect.

Moreover, tipping elements could also impact the GMT itself, i.e. the transitioning of an element into its new state generally involves positive or negative feedback to global warming \cite{mckay22}. The Earth's temperature is therefore dependent on the states of the tipping elements. Thus, if an element tips it could also affect the likelihood of other elements tipping by influencing the GMT.

\subsection{Global core tipping elements}
Tipping elements are classified into global core and regional tipping elements \cite{mckay22}. Global core tipping elements have a significant impact on the state of the Earth system as a whole, while regional tipping elements either have considerable influence on human welfare, or are a unique feature in the Earth system\cite{steffen18}. 
Tipping elements that have a significant impact (beyond $\pm 0.1$ °C) on the GMT and/or interact with other tipping elements are generally classified as global core tipping elements \cite{mckay22}.
We now briefly present six of the nine tipping elements, categorised as global core tipping elements\cite{mckay22}, that we consider in our network. 
Among these are three elements in the Earth's cryosphere, two ocean circulation elements and one biosphere element, see Table~\ref{tab:1}.
\begin{table}[htb]
\centering
\begin{tabular}{p{4.5cm} p{1.5cm} p{1.9cm}}
 \hline
Tipping element& $\Delta T_{\text{limit}}$ (°C)& Feedback (°C)\\
 \hline
 Greenland ice sheet (GIS)& 0.8-3.0& +0.13\\
 West Antarctic ice sheet (WAIS) & 1.0-3.0& +0.05\\
 Labrador-Irminger seas convection (LISC)& 1.1-3.8& -0.5\\
 East Antarctic subglacial basins (EASB)& 2.0-6.0& +0.05\\
 Amazon Rainforest (AR)& 2.0-6.0& +0.2\\
 AMOC& 1.4-8.0& -0.5\\
 \hline
\end{tabular}
    \caption{Range for the estimated critical threshold temperature and GMT feedback for each global core tipping element considered in the network, taken from Ref.~\onlinecite{mckay22}. With $"$GMT-feedback$"$ we mean the increase or decrease in temperature caused by the tipping of an element. Several of the estimated critical temperatures are updated compared to Ref.~\onlinecite{wunderling21}, based on more recent publications, the results of which are summarized in Ref. \onlinecite{mckay22}. 
    \label{tab:1}
    }
\end{table}

\paragraph{Greenland ice sheet (\rm{GIS})}
Greenland is covered by an ice sheet up to three meters thick, which could add about 7 meters to global sea level if completely melted \cite{lenton19, bochow23}. As the ice sheet melts, it loses height, causing its surface to be exposed to the warmer layers of air in lower altitudes, therefore accelerating the melting process. The Greenland ice sheet would then reach a tipping point once this so-called melt-elevation feedback induces self-perpetuating ice loss \cite{mckay22, levermann16}. If global temperatures are not reduced sufficiently quickly after an overshoot of the threshold, the ice loss would be irreversible \cite{bochow23}. Numerous studies suggest that the Greenland ice sheet is very close to, or may have already reached its tipping point \cite{mckay22, lenton19, boers21}.
\paragraph{West Antarctic ice sheet (\rm{WAIS})}
The West Antarctic ice sheet is mostly grounded on bedrock below sea level, causing it to be vulnerable to marine ice sheet instability due to ocean temperatures, which can lead to self-sustaining ice loss once the ice retreats beyond a certain point \cite{mckay22, joughin14, deconto21, schoof07}.\\
If tipped, the West Antarctic ice sheet would contribute about three meters of sea level rise \cite{lenton19}.
\paragraph{East Antarctic subglacial basins (\rm{EASB})}
Several subglacial basins in East Antarctica, most notably the Wilkes, Aurora, and Recovery basins, lie below sea level, meaning they are also vulnerable to marine ice sheet instability \cite{mckay22, lenton19}. The Wilkes basin alone could contribute further 3-4 meters of sea level rise \cite{lenton19}.
\paragraph{Atlantic meridional overturning circulation (\rm{AMOC})}
Driven by temperature and salinity anomalies, the AMOC transports warm surface water northwards, which increases its density due to cooling and evaporation \cite{mckay22}, causing it to sink. This cold deep water is then transported southwards. Global warming causes a rise in sea temperatures and freshwater influx due to melting of the Greenland ice sheet, which prevents deep water formation, therefore slowing down the circulation \cite{mckay22}. This slowdown is self-perpetuating, since northward salinity transport is reduced, which weakens the AMOC further \cite{levermann12}. An early warning indicator for the tipping of the AMOC was developed in Ref. \onlinecite{vandermheen13}.\\
An AMOC collapse would have far-reaching global impacts, including cooling in the North Atlantic, warming of the Southern hemisphere, a southward shift of the intertropical convergence zone, and reduced natural carbon sinks \cite{mckay22}.
\paragraph{Labrador-Irminger seas convection (\rm{LISC})}
As part of the North Atlantic subpolar gyre, there is an overturning circulation in the Labrador and Irminger seas as well, which collapses in several models due to global warming \cite{mckay22, ruehs21}. Such a collapse could affect the AMOC, as well as cause significant regional North Atlantic cooling and a southward shift of the intertropical convergence zone \cite{mckay22}.
\paragraph{Amazon rainforest (\rm{AR})}
The Amazon rainforest is an important sink for CO$^{2}$ emissions, storing up to 200 gigatonnes of carbon \cite{mckay22}. About a third of its precipitation is recycled by evaporation of water over the rainforest \cite{lenton08}. Global warming induced drying results in a dieback of the rainforest, which is self-perpetuating due to reduced precipitation \cite{mckay22}. Beyond a certain threshold this would result in the Amazon tipping into a dry savanna state or a degraded forest state \cite{mckay22}, \cite{flores24}. The resulting feedbacks would ensure that these transitioned states are stable \cite{flores24}. Amazon rainforest dieback is further reinforced by deforestation and forest fires \cite{wunderling21}, and deforestation beyond a certain percentage would also cause the Amazon to tip \cite{mckay22}. However, the critical temperature values given in Ref.~\onlinecite{mckay22} are independent of deforestation.

\subsection{Interactions}
 
The interaction links between two tipping elements $i$ and $j$ are quantitatively described in the model by a real number $s_{ij}$ with an absolute value between $1$ and $10$. The nature of the interaction is determined by the sign; if the sign is positive the tipping of element $j$ \textit{increases} the probability that element $i$ will tip as well, meaning the interaction is \textit{destabilizing}. A negative sign means that the interaction is \textit{stabilizing}; if element $j$ crosses its threshold, the probability that element $i$ will tip as well is decreased. The absolute value of $s_{ij}$ denotes the \textit{strength} of the interaction link, meaning how the probability for element $i$ crossing its tipping point is increased or decreased\cite{kriegler09,wunderling21}. For example, $s_{ij}=2$ means that if element $j$ tips, the probability that element $i$ tips as well is increased by a factor of $2$. We denote the maximum interaction link parameter for each link with $s_{ij}^{max}$.\\
In addition to the interaction links within the four-element network described in Ref.~\onlinecite{wunderling21}, we describe the interactions in an extended tipping element network which also includes the Labrador-Irminger seas convection and the East Antarctic subglacial basins. For a clearer notation we assign each tipping element with a number as follows: $\{\text{GIS},\text{WAIS},\text{AMOC},\text{AR},\text{EASB},\text{LISC}\}\longrightarrow \{1,2,3,4,5,6\}$.
\paragraph{GIS $\longleftrightarrow$ EASB.}
As mentioned in Ref. \onlinecite{wunderling21}, interactions between cryosphere elements are mutually destabilizing due to sea-level rise. However, the impact the collapse of the GIS has on the EASB should be larger than the impact of the interaction in the opposite direction since the EASB lie below sea-level, making them more vulnerable \cite{lenton19}. Since the EASB collapse has roughly the same contribution to sea-level rise as the WAIS collapse \cite{lenton19}, we set the corresponding interaction link parameter to $s_{1,5}^{max}=2$, the same as $s_{1,2}^{max}$. The parameter for the maximum interaction link in the opposite direction was cautiously set to $s_{5,1}^{max}=5$. However it is possible that a maximum parameter value of 10 like for the interaction between the GIS and WAIS would have been more sensible.
\paragraph{WAIS $\longleftrightarrow$ EASB.}
The WAIS and the EASB should mutually destabilize each other due to sea-level rise as well. Analogously to the maximum interaction link parameter values set for the interactions of these two elements with the GIS we set $s_{2, 5}^{max}=10$ and $s_{5,2}^{max}=5$. 
\paragraph{\rm{AMOC} $\longrightarrow$ EASB.}
According to Refs. \onlinecite{wunderling21}, \onlinecite{wunderling24} and \onlinecite{vanwesten24}, AMOC shutdown results in warming of the Southern hemisphere. Furthermore, this interaction link is explicitly stated as destabilizing in Ref. \onlinecite{lenton19}. Analogously to the interaction link strength between the AMOC and the WAIS, we therefore set the maximal link strength as $s_{5,3}^{max}=1.5$.
\paragraph{LISC $\longrightarrow$ GIS.}
According to Ref. \onlinecite{mckay22} a collapse would result in North Atlantic regional cooling of 2-3°C, which should have a stabilizing effect on the GIS. The maximal link strength is therefore set as $s_{1,6}^{max}=-10$.
\paragraph{LISC $\longrightarrow$ \rm{AMOC}.}
An interruption in the deep-water formation at the convection does weaken the AMOC, although deep-water formation has to collapse in all locations for an AMOC shutdown to occur \cite{sgubin17}. In the models discussed in Ref.~\onlinecite{sgubin17}, convection collapse is not associated with dramatic weakening of the AMOC even though deep-water formation is suspended, because this does not occur at all locations. We therefore classify the interaction as weakly destabilizing and set a low value of $s_{3,6}^{max}=1.5$ for the maximum link strength.
\paragraph{GIS $\longrightarrow$ LISC.}
In Ref. \onlinecite{sgubin17} the authors state that freshwater influx due to melting of the GIS is likely to increase the probability of a collapse, although this was not investigated. This is also suggestend in Ref. \onlinecite{ruehs21}. Due to lack of data, we set the interaction link as destabilizing, with a maximum value of $s_{6,1}^{max}=5$, although this value is somewhat arbitrarily chosen.
\paragraph{LISC $\longrightarrow$ AR.}
In Ref.~\onlinecite{jackson15}, cited by Ref.~\onlinecite{wunderling21}, to justify the unclear interaction link between the AMOC and the AR, it was stated that the southward shift of the Intertropical Convergence Zone induced by weakening of the AMOC causes changes in precipitation patterns over the Amazon.
As mentioned before, LISC collapse is also assumed to invoke a southward shift of the Intertropical Convergence Zone. Therefore, it is possible that the interaction link between the Labrador-Irminger seas convection and the Amazon should be classified as unclear as well. However, in Ref.~\onlinecite{swingedouw21} the authors classify the changes in precipitation caused by rapid regional cooling due to seas convection collapse as "modest" and mention that the climactic effects of LISC collapse are weaker than those of AMOC collapse. Furthermore, while it is mentioned in Ref.~\onlinecite{mckay22} that the effects of AMOC collapse include changes in precipitation patterns, including over the Amazon, such effects of LISC collapse were not mentioned. We therefore set this interaction link as zero, although it is possible that this interaction link should be set as unclear or weakly destabilizing.

\begin{table*}[htb]
    \centering 
    \caption{Maximum link strengths for each interaction link, including the subglacial basins in East Antarctica and the Labrador-Irminger seas convection. The link strengths of interactions within the original four-element network considered in Ref. \onlinecite{wunderling21} are taken from there. Interactions of the two additional elements are in \textbf{bold}. For the numerical study we consider interaction link strengths within the interval $[1,s_{ij}^{max}]$ for destabilizing interactions and $[s_{ij}^{max},-1]$ for stabilizing interactions.
    }
    \vspace{10mm} 
    \begin{tabular}{p{2.7cm} p{1.5cm} p{4cm} p{7cm}  }
Interaction link& & Maximum link strength $s_{ij}^{max}$ &Physical process\\
 \hline
 GIS $\rightarrow$ AMOC & (3,1)&  +10 & Freshwater influx\\
 AMOC $\rightarrow$ GIS & (1,3)&  -10 & Reduction in northward heat transport\\
 GIS $\rightarrow$ WAIS & (2,1)&  +10 &Sea level rise\\
 \textbf{LISC} $\mathbf{\rightarrow}$ \textbf{GIS} & \textbf{(1,6)}& 
  \textbf{-10} & \textbf{Regional cooling in the North Atlantic}\\
 \textbf{EASB} $\mathbf{\rightarrow}$ \textbf{WAIS} & \textbf{(2,5)}& 
  \textbf{+10} & \textbf{Sea level rise}\\
 \textbf{GIS} $\mathbf{\rightarrow}$ \textbf{EASB} & \textbf{(5,1)}& 
  \textbf{+5} & \textbf{Sea level rise}\\
 \textbf{WAIS} $\mathbf{\rightarrow}$ \textbf{EASB} & \textbf{(5,2)}& 
  \textbf{+5} & \textbf{Sea level rise}\\
 \textbf{GIS} $\mathbf{\rightarrow}$ \textbf{LISC} & \textbf{(6,1)}& 
  \textbf{+5} & \textbf{Freshwater influx}\\
 \hline
 AMOC $\rightarrow$ AR & (4,3)&  ±2 up to ±4 & Changes in precipitation patterns\\
 WAIS $\rightarrow$ AMOC & (3,2)&  ±3 & Increase in meridional salinity gradient (-), Fast advection of freshwater anomaly to North Atlantic (+)\\
\hline
 WAIS $\rightarrow$ GIS & (1,2)&  +2 & Sea level rise\\
\textbf{EASB} $\mathbf{\rightarrow}$ \textbf{GIS} & \textbf{(1,5)}& \textbf{+2} & \textbf{Sea level rise}\\
 AMOC $\rightarrow$ WAIS & (2,3)&  +1.5 & Heat accumulation in Southern Ocean\\
 \textbf{LISC} $\mathbf{\rightarrow}$ \textbf{AMOC} & \textbf{(3,6)}& \textbf{+1.5} & \textbf{Suspension of local deep-water formation}\\
 \textbf{AMOC} $\mathbf{\rightarrow}$ \textbf{EASB} & \textbf{(5,3)}& \textbf{+1.5} & \textbf{Heat accumulation in Southern Ocean}\\
 \textbf{LISC} $\mathbf{\rightarrow}$ \textbf{AR} & (4,6)& $\mathbf{\pm0}$\textbf{ (?)} & \textbf{Changes in precipitation patterns. Interaction set to 0. (?) denotes high uncertainty}
	\end{tabular}
 \label{tab:2}
\end{table*}

\section{Model}
We now present a conceptual model to describe basic phenomena and interactions within the network of tipping elements.
\subsection{Dynamics of a single tipping element}
\begin{figure}[htb!]
    \centering
    \includegraphics[scale=0.5]{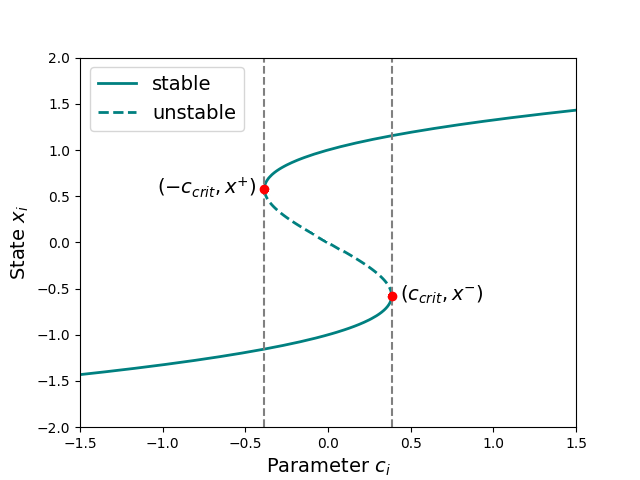}
    \caption{Bifurcation diagram for a double-fold bifurcation in system \eqref{eq:single-element}. 
    }
    \label{fig:mesh2}
\end{figure}

The state $x_{i}$ of a single tipping element $i$, without considering  any interactions with other tipping elements, is modeled in Ref.~\onlinecite{wunderling21} by the following excitable nonlinear system
\begin{equation}
\label{eq:single-element}
\tau_{i}\cdot\dfrac{dx_{i}}{dt}= -x_{i}^{3}+x_{i}+c_{i},
\end{equation}
with a timescale $\tau_i$. This system exhibits two coexisting stable states and a hysteresis, see Fig.~\ref{fig:mesh2}.

As described in Refs. \onlinecite{wunderling21}, \onlinecite{boers22} and \onlinecite{klose20}, this type of bifurcation is suitable for describing the behavior of a tipping element with two stable states. These correspond to its baseline, pre-industrial state and the transitioned state it would eventually reach if tipped. The stable baseline state is set as $x_{i}\approx -1$ (lower branch) and the stable transitioned state is set as $x_{i}\approx 1$ (upper branch). We vary the critical parameter $c_{i}$, which denotes the temperature rise, starting from the baseline state. This is the only equilibrium until the critical value $c_{i}=-c_{\text{crit}}=-\sqrt{\frac{4}{27}}$ is reached, where another equilibrium appears, namely the transitioned state. By increasing $c_{i}$ further to the critical value $c_{i}=c_{\text{crit}}$, the equilibrium corresponding to the stable baseline state disappears and only the stable transition state is left. Due to hysteresis behavior, this transition cannot simply be reversed by slightly lowering $c_{i}$ below $c_{\text{crit}}$. The thresholds for the baseline and the transitioned states are $x^{-}=-\frac{1}{\sqrt{3}}$ and $x^{+}=\frac{1}{\sqrt{3}}$, respectively, i.e. a tipping element is considered tipped once its state reaches $x=x^{+}$.

\subsection{Modelling the network}
The interactions between the tipping elements are incorporated in the model by extending \eqref{eq:single-element} by a linear coupling term:
\begin{equation}
\tau_{i}\cdot\frac{dx_{i}}{dt}=-x_{i}^{3}+x_{i}+c_{i}+\frac{1}{2}\sum_{j\neq i}d_{ij}(x_{j}+1),
\end{equation}
yielding a weighted dynamical network with the same coupling function for all links. A tipping element can now not only tip into the transitioned state due to its critical temperature threshold being crossed, but also by interacting with other tipping elements.

We parameterize the system of equations by setting $c_{i}=c_{\text{crit}} \cdot \frac{\Delta \mathrm{GMT}}{T_{limit, i}}$ and $d_{ij}=d\cdot \frac{{s_{ij}}}{5}$ (see Ref.~\onlinecite{wunderling21}), where $\Delta \mathrm{GMT}$ is a parameter denoting the rise in GMT. This way, a tipping element transgresses into the transitioned regime as soon as $\Delta \mathrm{GMT} > T_{limit,i}$, i.e. its critical temperature is exceeded. This yields the following parameterized system of equations:
\begin{equation}
\label{eq:nonadaptive}
\tau_{i}\cdot\dfrac{dx_{i}}{dt}=-x_{i}^{3}+x_{i}+\sqrt{\dfrac{4}{27}}\cdot \dfrac{\Delta \mathrm{GMT}}{T_{limit, i}}+d\sum_{j\neq i}\dfrac{s_{ij}}{10}(x_{j}+1).
\end{equation}
Now $d$ denotes the overall interaction strength between the tipping elements, which is varied in the experiments, and $s_{ij}$ denotes the link strength between the tipping elements $i$ and $j$. The proposed values for $s_{ij}$ are summarized in Table~\ref{tab:2} and discussed in Section II.C.

\subsection{Adaptation}

As described in Section II, a tipping element crossing its threshold may impact the global mean surface temperature, i.e., the GMT itself depends on the states of the tipping elements. We therefore extend the system describing the dynamics within the network of tipping elements to an \textit{adaptive} system, 
where the GMT depends on the states $x_{i}$ of the tipping elements. The system from Ref.~\onlinecite{wunderling21} described above corresponds to the case where the GMT is considered as a constant parameter, meaning it is non-adaptive.

In our model, if an element tips into the transitioned state, it can either induce an increase or a decrease in the GMT. Since a tipping element is considered tipped once its state reaches $x_{i}=x^+=\frac{1}{\sqrt{3}}$ the adaptive mechanism should incorporate a continuous rise by a certain feedback temperature value (stated for each element in Table~\ref{tab:1}), starting when $x_{i}$ reaches $x^+$ and stopping when $x_{i}=1$, i.e. the element is completely tipped.

For the adaptive mechanism, we propose for $\Delta \mathrm{GMT}$ a function of the form:
\begin{align}
\Delta \mathrm{GMT} = \Delta \mathrm{GMT}_{0} + \sum_{i} \eta_{i} g(x_{i}-x^+),
\end{align}
where $\Delta \mathrm{GMT}_{0}$ is a baseline value for the GMT rise in the model, $\eta_{i}$ a parameter, which varies for each tipping element, and $g$ an activation function. Only when the state $x_{i}$ of a certain tipping reaches the value $x^+$, it contributes to the GMT.

\begin{figure}[htb!]
    \centering
    \includegraphics[scale=0.5]{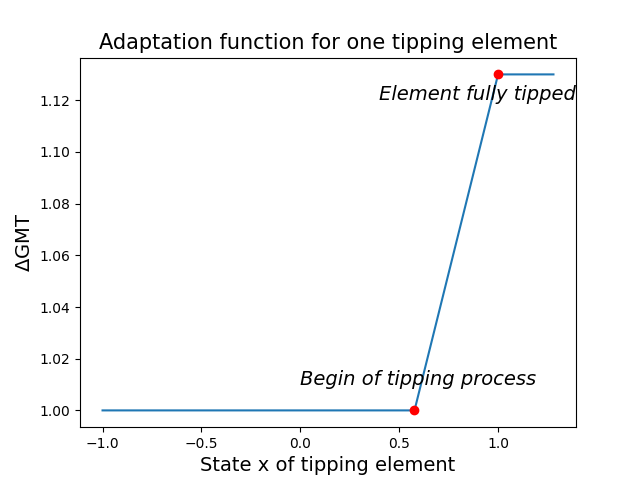}
    \caption{Visualization of the adaptation function ($\Delta\text{GMT}(x)$ in \eqref{eq:adaptive}) for a single tipping element (the GIS was taken as an example) and baseline $\Delta\text{GMT}_0$ value of 1.0°C. Once the element crosses its critical threshold, i.e. its state reaches $x_i = \frac{1}{\sqrt{3}}$, the GMT increases linearly, until the element's GMT-feedback (in this case 0.13°C) is reached when the element is completely tipped.
    }
    \label{fig:mesh6}
\end{figure}

For the activation function $g$ we use the scaled ReLU-function defined by
\begin{align*}
f_{ReLU,\alpha}(x)=\max(0,\alpha x).
\end{align*}
However, the states $x_{i}$ of the tipping elements usually converge to a value larger than one, which would result in further warming in the model. Therefore, we adapt the ReLU-function to construct the following activation function:
\begin{align}
\label{eq:g}
g(x):=
\begin{cases}
f_{ReLU,\alpha}(x), & \text{if}\ x < 1 - x^+\\
1 , & \text{otherwise}\
\end{cases}
\end{align}
The parameter $\alpha$ is set as $\frac{1}{1-x^+}$, so that when the value of $g(x_{i}-x^+)$ reaches 1 when the state of the element reaches $x_{i}=1$, i.e. the element is fully tipped, see Fig.~\ref{fig:mesh6}. For the parameters $\eta_{i}$, we set for each element its proposed feedback to the GMT, if tipped. This way, once the activation function reaches the value 1, the feedback to the GMT is fully contributed.

The final model with adaptation is therefore
\begin{equation}
\label{eq:adaptive}
\begin{split}
    &\tau_{i} \cdot \dfrac{dx_{i}}{dt}=-x_{i}^{3}+x_{i}+\sqrt{\dfrac{4}{27}}\cdot \dfrac{\Delta \mathrm{GMT}(x)}{T_{limit, i}}+d\sum_{j\neq i}\dfrac{s_{ij}}{10}(x_{j}+1),\\
    &\Delta \mathrm{GMT}(x) = \Delta \mathrm{GMT}_{0} + \sum_{i} \eta_{i} g(x_{i}- x^+),
\end{split}
\end{equation}
with $x_{i}$ the state of the tipping element $i$, $T_{limit,i}$ the critical temperature of tipping element $i$, $\tau_{i}$ the tipping timescale of tipping element $i$, $s_{ij}$ the interaction link strength between the tipping elements $j$ and $i$, $\eta_{i}$ the GMT-feedback caused by tipping of tipping element $i$, and $g(\cdot)$ the activation function defined by equation \eqref{eq:g}. $\Delta \mathrm{GMT}_{0}$ denotes the baseline value for the GMT-rise and $d$ is the overall interaction strength.

It should be noted that this model does not have the form of an adaptive or coevolutionary dynamical network \cite{yanchuk23}, since the coupling weights remain constant. The state of the tipping elements in the network, therefore, does not influence the topology of the network, i.e. any change in the state does not result in any links appearing or disappearing, or changing their weights. However, the model could be regarded as belonging to a more general class of adaptive dynamical systems\cite{sawickiPerspectivesAdaptiveDynamical2023}, since the GMT rise determines whether any tipping elements transgress their thresholds, but is now also dependent on the states of the tipping elements themselves.

A tipping element can now also tip into the transitioned state due to further global warming induced by the tipping of another element. This means that a tipping element can not only initiate cascades by having outgoing destabilizing interaction links with other elements, but also by having a positive feedback on the GMT. 
Due to its +0.2 ° C feedback to the GMT, the Amazon rainforest can initiate cascades in the adaptive model even though it has no outgoing interaction links in the model.
This will be confirmed numerically below. 

\section{Numerical study}
We now repeat the experiments conducted in Ref.~\onlinecite{wunderling21} for the extended network of six tipping elements with adaptation (equation \eqref{eq:adaptive}). In order to assess effects of the adaptation mechanism, we also consider the same network without adaptation and compare the results for both networks in section~\ref{sec:V}. The procedures used for the numerical study are mostly the same as in Ref.~\onlinecite{wunderling21}, but for the sake of completeness we will recall them here.
\paragraph*{General experimental setup}

(i) For each pair $(\Delta \mathrm{GMT}, d)$ of $\Delta \mathrm{GMT} \in [0,8]$°C (step-size 0.1°C) and interaction strength $d \in [0,1]$ (step-size 0.02), 100 values for each critical temperature $T_{limit,i}$ and $s_{ij}$ are randomly drawn independently of each other from a uniform probability distribution over their respective uncertainty ranges (intervals given in Tables \ref{tab:1}, \ref{tab:2}). Since we have two uncertain interactions (WAIS $\rightarrow$ AMOC, AMOC $\rightarrow$ AR), we use the same 100 sets of parameter values generated by the Monte-Carlo sampling for each permutation (i.e. WAIS $\rightarrow$ AMOC destabilizing and AMOC $\rightarrow$ Amazon stabilizing, both interactions destabilizing, both interactions zero, etc.).

(ii) The initial conditions for the system are set as follows: the state $x_i,\; i=1,...,6$ of each element is set as $x_i = -1$, corresponding to the stable pre-industrial baseline state of the tipping elements. 

(iii) We now run the simulation for the time interval $[0,200000]$. As in Ref.~\onlinecite{wunderling21}, the simulation time is chosen to be significantly larger than the assumed tipping timescale of the Greenland ice sheet. As stated in the previous section, an element is considered tipped once its state $x_i$ reaches the value $\frac{1}{\sqrt{3}}$. A tipping cascade (or tipping event if only one element has transitioned) is defined to have occurred for a fixed pair $(d,\Delta \mathrm{GMT})$ if there are elements that cross the threshold for $(d,\Delta \mathrm{GMT})$, but not for $(d,\Delta \mathrm{GMT}-0.1)$. The number of these elements defines the size of the cascade.

\begin{figure*}[htb]
    \subfigure[]
    {
    \includegraphics[scale=0.4]{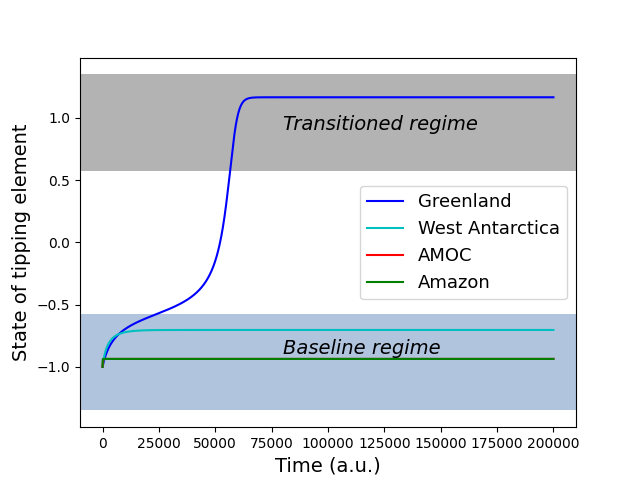}
    }
    \subfigure[]
    {
    \includegraphics[scale=0.4]{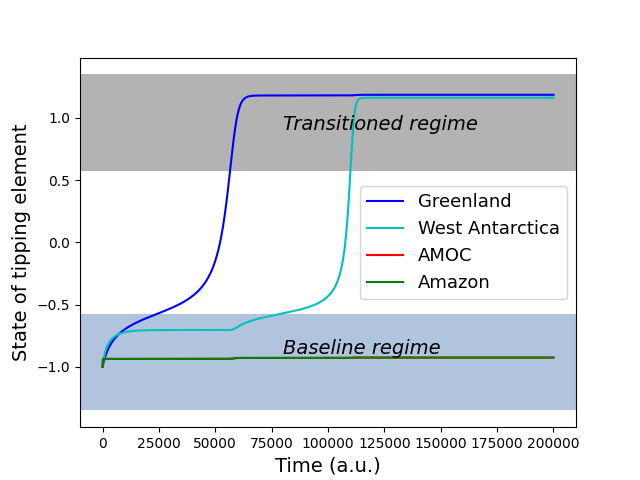}
    }
    \subfigure[]
    {
    \includegraphics[scale=0.4]{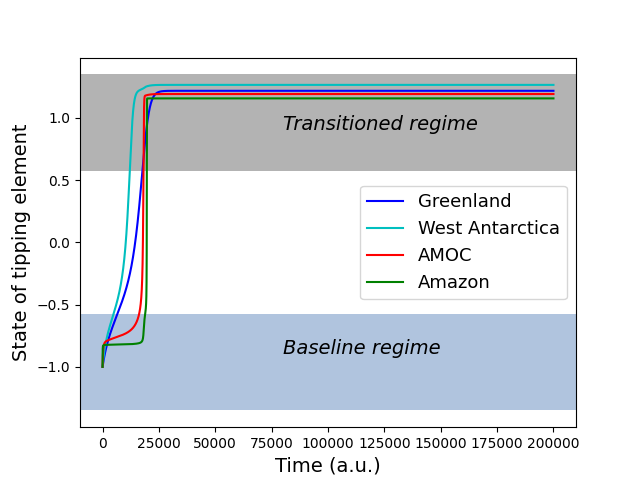}
    }
    \subfigure[]
    {
    \includegraphics[scale=0.4]{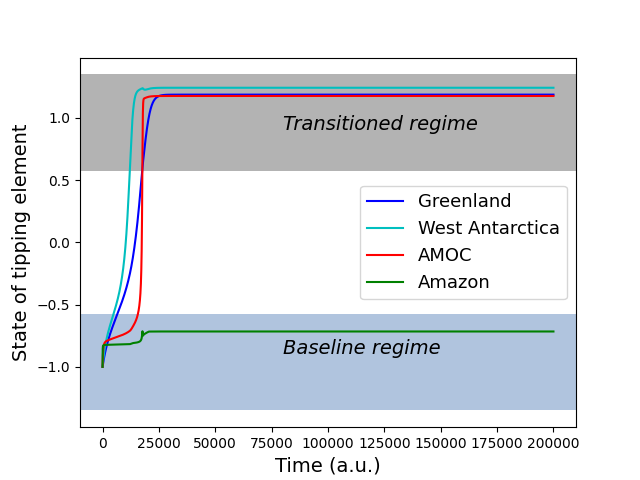}
    }
    \caption{(a)-(d): Exemplary time series to visualize the two modes of operation of the adaptive model, using a network of only four tipping elements for clarity. The time is measured in years, this is, however, not relevant here since the model may not be suitable for projecting how long a cascade would take to unfold \cite{wunderling21}. Figure (a) depicts a single tipping event occurring in the non-adaptive model, in which only the Greenland ice sheet tips into its transitioned state. The same choice of parameter values yields however a two-element tipping cascade in the adaptive model, where the additional global warming due to tipping of the GIS causes the WAIS to tip as well (Figure (b)). A different choice of parameter values for the non-adaptive model yields the four-element tipping cascade depicted in Figure (c). By contrast, setting the same parameter values into the adaptive model yields only a three-element cascade, as depicted in Figure (d), due to the AMOC's negative feedback to the GMT preventing the Amazon from tipping.
    }
    \label{fig:mesh3}
\end{figure*}

We conduct two separate experiments:
\paragraph{Tipping within GMT rise of 2°C.} In the first experiment, whose results are summarized in Table~\ref{tab:3}, we examine the probability for the occurrence of tipping cascades for GMT rise up to 2°C. Here, for each interaction strength $d\in [0,1]$, 100 sets of parameter values are randomly chosen. For each set, we run the simulation, starting with GMT-value 0. As long as a tipping event (cascade) is not recorded, we raise the GMT by 0.1°C and solve the system again. If a tipping event (cascade) is recorded for GMT up to 2°C, we count it as a cascade of its respective size and move on to the next set of parameter values. If no tipping event is recorded, the ensemble member does not contain a cascade within the 2°C limit.
An exemplary time series of the adaptive model is shown in Fig.~\ref{fig:mesh3}.

\paragraph{Occurrence of cascades and role of each tipping element.}
Here, we look at the distribution of cascades over values for GMT-rise between 0 and 8°C.  Furthermore, in order to assess the roles of the individual tipping elements in cascades, we count the number of times an element occurs in a cascade and the number of times it is an initiator of cascades. As in Ref.~\onlinecite{wunderling21}, the \emph{initiator} of a cascade is defined as the element whose critical threshold temperature is closest to the $\Delta$GMT-value for which the tipping occurred.

It should be noted that for the adaptive model the range of $\Delta \mathrm{GMT}$-values used now defines the baseline GMT-value $\Delta \mathrm{GMT}_{0}$. The graphs are, therefore, plotted with respect to $\Delta \mathrm{GMT}_{0}$ and not with respect to the temperature for which the tipping actually occurred.

For better comparability, the same set of parameter values is used for both the adaptive and non-adaptive models.
\paragraph*{Adjustments to the original model from Ref.~\onlinecite{wunderling21}.} Since Ref. \onlinecite{wunderling21} was published, the boundaries for the estimated critical temperature thresholds of the West Antarctic ice sheet, the AMOC and the Amazon rainforest were adjusted \cite{mckay22}. The numerical study in this paper is, therefore, conducted using the updated values, see Table~\ref{tab:1}. 

\subsection{Risk of tipping cascades}
\begin{table*}[htb!]
    \centering 
    \caption{Share of tipping events in ensemble simulation in the model with adaptation for values of the interaction strength $d$ up to a maximum value of $d_\text{max}$. The results for the model without adaptation are written in brackets for reference.}
    \vspace{10mm} 
	\begin{tabular}{p{1.7 cm} p{2 cm} p{2 cm} p{1.5 cm} p{1.5 cm} p{1.5 cm} p{1.5 cm} p{1.5 cm} p{1.5 cm}}
	\hline 
	Maximum& No& Tipping& \multicolumn{6}{c}{Cascade sizes (\%)}\\
	\cline{4-9}\\
	interaction strength $d_\text{max}$& tipping (\%)& (\%)& 1& 2& 3& 4& 5& 6\\
	\hline
	1.0& 5 (5)& 95 (95)& 32 (31)& 8 (7)& 29 (27)& 19 (20)& 6 (9)& 1 (1)\\
	0.75& 6 (6)& 94 (94)& 39 (38)& 10 (9)& 22 (22)& 17 (17)& 5 (7)& 1 (1)\\
	0.5& 8 (8)& 92 (92)& 47 (47)& 12 (11)& 15 (14)& 14 (14)& 3 (5)& 0 (1)\\
	0.25& 11 (11)& 89 (89)& 60 (62)& 15 (14)& 8 (6)& 5 (5)& 1 (1)& 0 (0)\\
	0.1& 15 (15)& 85 (85)& 72 (75)& 12 (9)& 1 (1)& 0 (0)& 0 (0)& 0 (0)\\
	\hline
	\end{tabular}
 \label{tab:3}
\end{table*}
As seen in Table~\ref{tab:3}, 95\% of ensemble simulations contain a tipping event for GMT rise up to 2°C if all values for the interaction strength $d$ are considered. Most ensemble simulations contain either a one-element tipping event (32\%), or three- or four-element tipping cascades (29\% and 19\%, respectively). Two-element cascades are relatively uncommon (8\%) due to the high number of destabilizing interaction links within the model.

Tipping cascades are first induced for a GMT-rise of about 0.5-0.6 °C (Figure~\ref{fig:mesh4}), which is slightly lower than the lower bound of the temperature threshold of the most vulnerable element, the GIS. This holds for all cascade sizes. Most tipping cascades consisting of at least three elements occur within the limit of 2°C, however, two-element cascades occur frequently for higher temperatures as well, since three of the tipping elements are expected to have relatively high critical temperature thresholds.

\subsection{Roles of tipping elements in cascades}
In Ref.~\onlinecite{wunderling21}, the authors assign each element in the original four-element network a role in tipping cascades, with the initiator being the element triggering a cascade. The last element to occur in a cascade is called a follower, and elements tipping before the follower are called mediators of cascades. It should be noted that the initiator of a tipping cascade is defined in Ref.~\onlinecite{wunderling21} as the element whose critical temperature threshold is closest to the GMT rise for which the cascade occurs, which does not mean that the initiator itself necessarily has to tip. This can be the case for the GIS due to its ingoing stabilizing interaction links.

\begin{table}[htb!]
\centering 
    \caption{Occurrence of each tipping element in cascades, percentage of cascades initiated by each element in the non-adaptive model.}
    \vspace{10mm} 
\begin{tabular}{p{3.5cm} p{2.5cm} p{2.5cm} }
 \hline
Tipping element& Initiator of cascades (\%)& Occurrence in cascades (\%)\\
 \hline
 Greenland ice sheet& 45& 37\\
 West Antarctic ice sheet& 37& 87\\
 AMOC& 5& 68\\
 Amazon rainforest& 0& 16\\
 Subglacial basins& 3& 88\\
 Labrador-Irminger& 11& 30\\ 
 \hline
\end{tabular}\\
\label{tab:4}
\end{table}
\begin{table}[htb!]
\centering 
    \caption{Occurrence of each tipping element in cascades, percentage of cascades which each element initiates in the adaptive model.}
    \vspace{10mm} 
\begin{tabular}{p{3.5cm} p{2.5cm} p{2.5cm} }
 \hline
Tipping element& Initiator of cascades (\%)& Occurrence in cascades (\%)\\
 \hline
 Greenland ice sheet& 42& 36\\
 West Antarctic ice sheet& 36& 87\\
 AMOC& 4& 61\\
 Amazon rainforest& 5& 15\\
 Subglacial basins& 4& 85\\
 Labrador-Irminger& 7& 29\\ 
 \hline
\end{tabular}\\
\label{tab:5}
\end{table}

Most tipping cascades are triggered by collapse of one of the ice sheets (Table~\ref{tab:5}). The other four elements are initiators of 4-7\% of cascades each. The tipping elements occurring most frequently in cascades are the WAIS and the EASB (87\% and 85\%, respectively), followed by the AMOC (61\%). All these elements have a high number of destabilizing ingoing interaction links, the AMOC, however, also has one unclear ingoing interaction link. This, together with its negative GMT-feedback, causes it to occur less frequently than the two Antarctic tipping elements. The three other tipping elements occur significantly less frequently in cascades, since they either have a low number of destabilizing interaction links (LISC, Amazon), or have stabilizing ingoing interaction links (GIS). 

\section{Comparison \label{sec:V}}
\subsection{Risk of tipping cascades}
\begin{figure*}[htb!]
\includegraphics[scale=0.4]{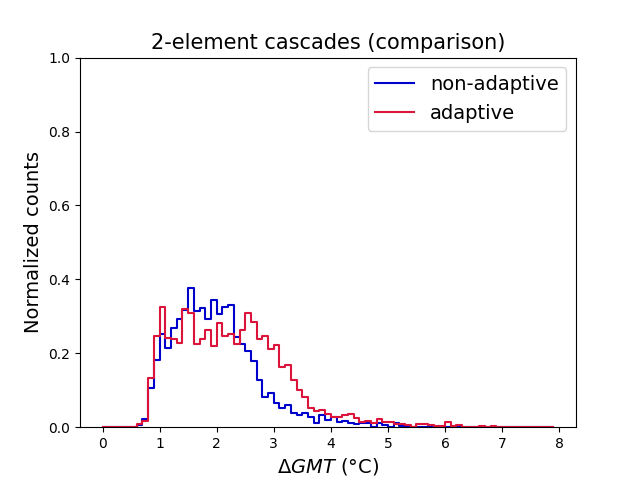}
\includegraphics[scale=0.4]{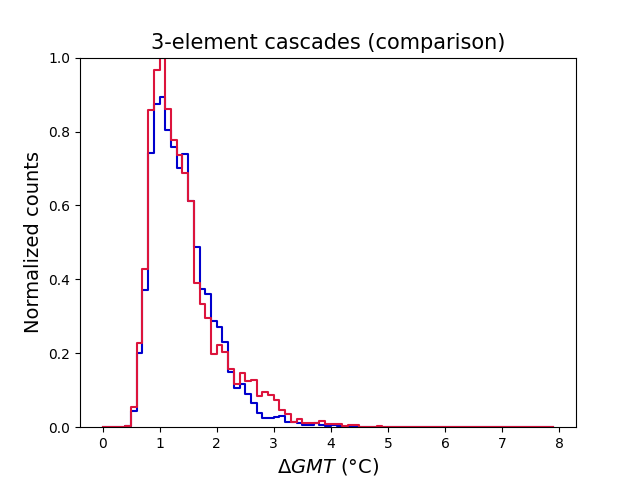}
\includegraphics[scale=0.4]{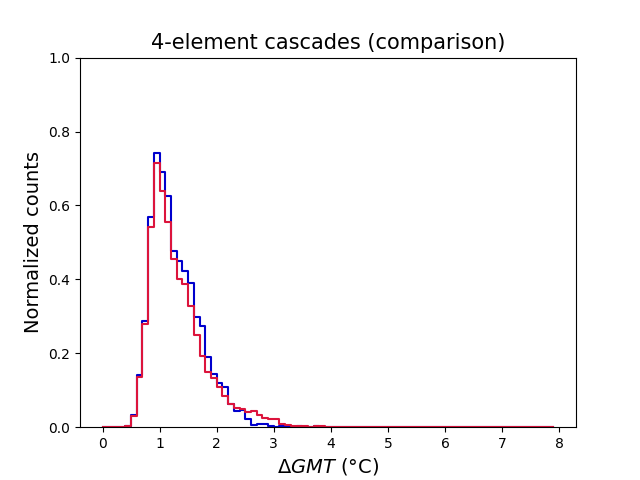}
\includegraphics[scale=0.4]{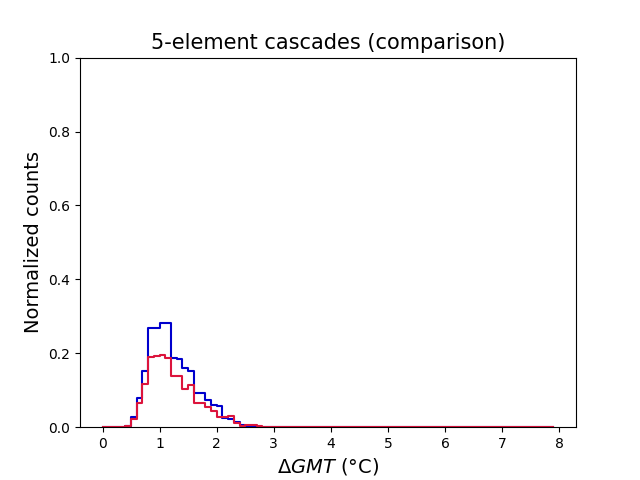}
\includegraphics[scale=0.4]{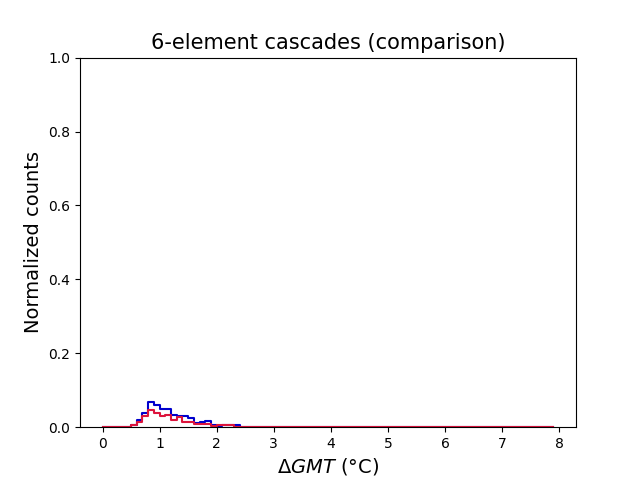}
 \caption{Occurrence of two-, three-, four-, five-, and six-element cascades for both the model with adaptation \eqref{eq:adaptive} (red) and without adaptation \eqref{eq:nonadaptive} (blue). All counts are normalized with respect to the highest value of three-element cascades in the adaptive model.}
 \label{fig:mesh4}
\end{figure*}
\begin{figure*}[htb!]
\includegraphics[scale=0.4]{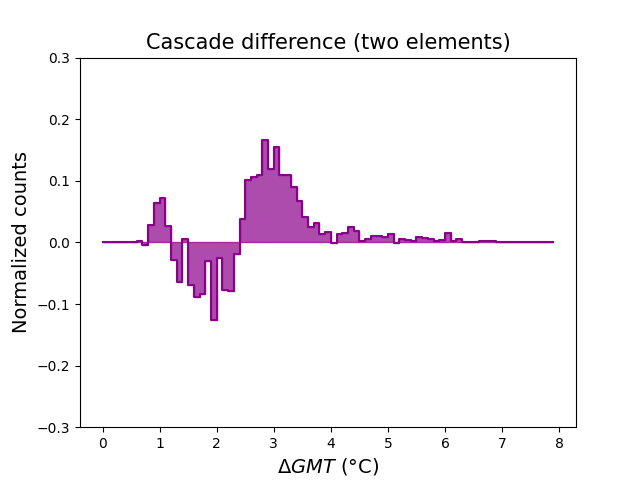}
\includegraphics[scale=0.4]{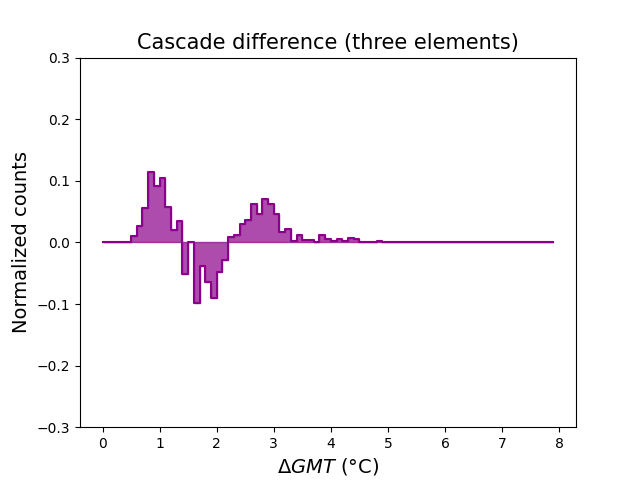}
\includegraphics[scale=0.4]{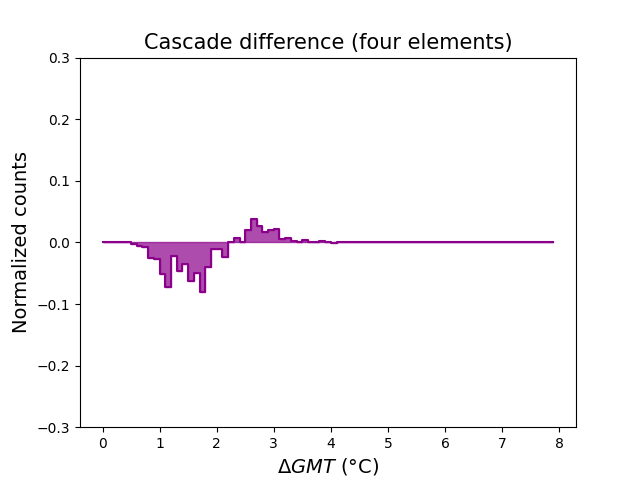}
\includegraphics[scale=0.4]{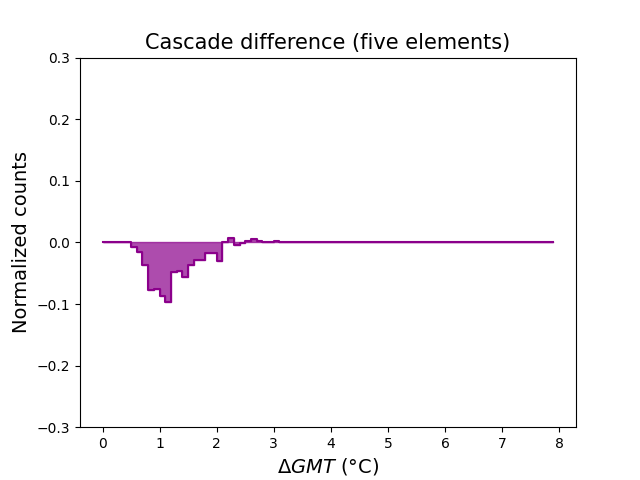}
\includegraphics[scale=0.4]{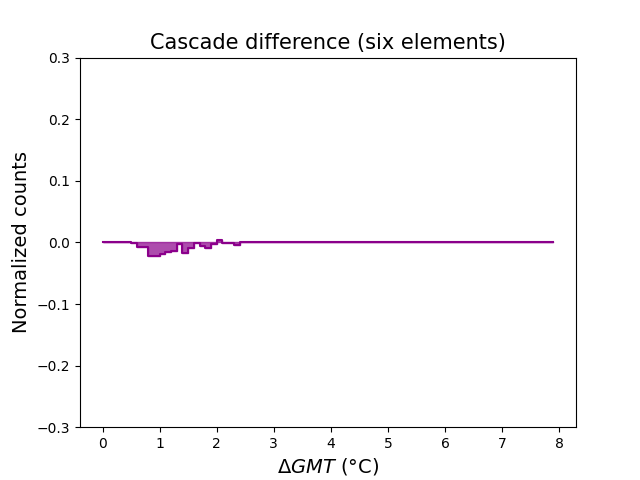}
 \caption{Difference (i.e. normalized counts for the adaptive model minus the normalized counts for the non-adaptive model) between the number of two-, three-, four-, five-, and six- element cascades occurring in the adaptive model and the non-adaptive model.}
 \label{fig:mesh5}
\end{figure*}

As seen in Table~\ref{tab:3}, the risk for the occurrence of tipping events is not influenced by the adaptation. This is due to the fact that in our adaptive model, a tipping element has to cross its critical threshold for it to influence the GMT. However, there is a slight difference in cascade sizes, with larger cascades being slightly less probable for the adaptive model if all values for the interaction strength $d$ are considered. Interestingly, for small values of $d$ ($<0.25$) the adaptation mechanism is slightly destabilizing, with cascades of at least two elements occurring more frequently in the adaptive model. The explanation for this is the following: for low values for $d$, whether an element has tipped or not is mostly dependent on its critical temperature threshold, and not on the interactions. The elements most likely to reach a tipping point first are the two ice sheets, which have a positive GMT-feedback, i.e. for some parameter values the rise in temperature causes another element to be part of a cascade. In contrast, high values for $d$ induce larger cascades due to the interactions between the tipping elements. Some of these elements, however, have a negative GMT-feedback, preventing further elements from occurring in a cascade.

In Figure~\ref{fig:mesh4}, we compare the occurrence of tipping cascades of each size in the adaptive and non-adaptive models. We see that in the adaptive model, tipping cascades occur in general for a wider range of temperatures and are more evenly distributed over the GMT values, i.e., they occur for higher temperatures than in the model without adaptation. This is mainly true for cascade sizes between two and four; for five- and six-element cascades the GMT-range in which cascades occur remains roughly the same for both models. There are two possible explanations for this. Either (a), the positive GMT-feedback of a tipping element induces a two-element cascade for parameter values for which there would have only been a one-element tipping event in the non-adaptive model, or (b), the negative GMT-feedback of a tipping element causes several cascades to only be able to occur for higher baseline GMT values compensating the decrease in GMT. However, as seen in Table~\ref{tab:1}, the GMT decrease expected to be induced by collapse of the circulation systems is significantly larger than the GMT increase induced by the tipping the other elements. On account of this and further results described below, we conclude that the latter is much more often the case.

For a better comparison, we plot the difference between the occurrence of tipping cascades in the adaptive and non-adaptive models in Figure~\ref{fig:mesh5}. More two- and three-element cascades occur in the adaptive model, however, the occurrence of larger cascades, especially of five and six elements, is reduced. This can, again, mostly be explained by the negative GMT-feedback that both circulation systems have, meaning that their tipping lowers the probability of further elements occurring in a cascade. The difference between both models is not evenly distributed over the range of GMT-values. We see that two- and three-element cascades occur more frequently in the adaptive model for GMT-rise up to about 1.2°C, significantly less frequently for GMT-rise up to 2.5°C, and more frequently again for higher temperatures. It is, however, clear that overall, both cascades sizes occur much more frequently in the adaptive model.

\subsection{Roles of tipping elements in cascades}
In Tables \ref{tab:4} and \ref{tab:5} we compare the percentage of cascades triggered by each tipping element, and the percentage of cascades in which each element occurs, in the non-adaptive and adaptive model, respectively. In both the adaptive and non-adaptive models, the overwhelming majority of cascades are caused by the tipping of one of the ice sheets ($\sim$80\%). Both ice sheets can therefore be considered as \emph{initiators} of cascades, regardless of the adaptation. Most of the tipping elements occur in roughly the same percentage of cascades in both models, albeit the percentages are slightly lower for the adaptive model. This can probably be attributed to the results described in Section A; smaller cascades in the adaptive model mean that each tipping element occurs in them with smaller probability. The exception is the AMOC, which occurs in significantly less cascades in the adaptive model. This can probably be attributed to the negative GMT feedback of the AMOC, even though we do not see the adaptation having a similar effect on the LISC, which has a similar negative feedback on the GMT. For the LISC the negative GMT-feedback is probably compensated by its low critical temperature threshold. Overall, the changes in percentages are mostly minor ($<5$\%), indicating that the original non-adaptive model is relatively robust.

We conclude that the occurrence of a tipping element in cascades is mostly dependent on its destabilizing ingoing interaction links, although a strong GMT feedback has a contribution as well, as seen in the case of the AMOC.

The inclusion of the adaptation into the model can cause several tipping elements to alter their role in tipping cascades, compared to the model without adaptation. Since the Amazon is a pure follower in the non-adaptive model, it cannot initiate any cascades. In the adaptive model, however, it can initiate cascades due to its feedback to the GMT.

Due to the negative GMT-feedback of both circulation systems, their role in cascades can be altered from mediators to followers in the adaptive model.

\section{Conclusions}
We find that the inclusion of more tipping elements as well as the usage of recent estimates for the critical temperature thresholds lead to a further destabilization of the system, compared to the results in Ref.~\onlinecite{wunderling21}. The additionally considered adaptation has a slightly stabilizing effect because the presumed negative GMT-feedback of the circulation systems is much larger than the positive GMT-feedback of the other elements. 
The probability for the occurrence of large tipping cascades is, therefore, reduced when adaptation is taken into consideration. Furthermore, we see that while the original four elements  considered in Ref.~\onlinecite{wunderling21} retain their roles in tipping cascades in the extended, non-adaptive six-element network, the adaptation can cause several elements to play qualitatively different roles in cascades.

Note, however, that due to the lack of a rigorous study of the interactions related to the two additional tipping elements, as was done for the original four-element network in Ref.~\onlinecite{kriegler09}, some interaction link strengths may be adjusted in future studies. 
It should also be noted that the interaction links between the tipping elements considered in this paper may be incomplete; for example, in Ref.~\onlinecite{kurths23} the authors find negative teleconnections linking the Amazon rainforest to the WAIS. Consequently, the Amazon may play a more active role in tipping cascades than what this study is considering. Furthermore, the values for the GMT feedback of each element were set as constants, even though there is uncertainty with respect to the exact impact \cite{mckay22}. It is therefore possible that incorporating these uncertainties into the model would have been more appropriate. The main goal of this paper is, however, to propose an adaptive model that can serve as a basis for future studies after some of the uncertainties, mostly regarding the interactions between tipping elements, become clearer.

In this study, as well as in Ref.~\onlinecite{wunderling21}, the coupling link weights as well as the temperatures in the model are considered as parameters. An approach in which the system is modeled as an adaptive dynamical network, in which e.g. the link strengths depend on the states of the tipping elements, or on one another as well, yields a more accurate description of the climate system. In Ref.~\onlinecite{yanchuk23} the authors proposed that the AMOC can be seen as an adaptive coupling between the GIS and WAIS due to its role as a mediator transmitting cascades, meaning that the coupling link weights between these two elements can vary, depending on the state of the AMOC. Following this idea, it may be possible to consider as an adaptive coupling e.g. the El Niño–Southern Oscillation (ENSO), which may have an impact on the states of the tipping elements \cite{kriegler09, wunderling21, wunderling24, fan21, singh20}. A further simplification is that the physical processes themselves are not considered in the model \cite{sinet23}. Some interactions are actually dependent on the derivatives of the states, rather than on the states of the tipping elements themselves.

Finally, we note that model \eqref{eq:adaptive} does not include the inertia for the change of $\Delta$GMT, and a possible extension of the model would be to include the first derivative of $\Delta$GMT in the GMT adaptation equation, so that the change in $\Delta$GMT is not instantaneous, but with some lag, which is probably a more realistic approach. 
\begin{acknowledgments}

\end{acknowledgments}
This work was supported by the German Research Foundation DFG, Project No. 411803875.

\appendix

\bibliography{aipsamp}

\end{document}